# Evidence for high inter-granular current flow in single-phase polycrystalline MgB$_2$ superconductor

K.Kawano*, J.S.Abell*, M.Kambara**, N Hari Babu** and D.A.Cardwell**
*School of Metallurgy and Materials, University of Birmingham, Birmingham B15 2TT, UK
**IRC in Superconductivity, University of Cambridge, Madingley Road, Cambridge CB3 0HE, UK

(April 6, 2001)

***Generalised abstract*** : The distribution of magnetic field in single-phase polycrystalline bulk MgB$_2$ has been measured using a Magneto-Optical (MO) technique for an external magnetic field applied perpendicular to the sample surface. The MO studies indicate that an inter-granular current network is readily established in this material and the current is not limited by weak-linked grain boundaries. The grain boundaries are observed to resist preferential magnetic field penetration, with the inter-grain mechanism dominating the current flow in the sample at temperatures up to 30K. The results provide clear evidence that the intra-granular current flow is isotropic. A critical current density of $\approx 10^4$ Acm$^{-2}$ was estimated at 30K in a field of 150mT from the MO measurements. These results provide further evidence of the considerable potential for MgB$_2$ for engineering applications.

The observation of superconductivity in MgB$_2$ at the remarkably high transition temperature of $\approx$ 39K [1] has aroused considerable interest in the current carrying properties of this material for engineering applications. Inevitably, comparisons have been drawn between the properties of this material and those of the established high temperature superconductors (HTS). In the case of the latter, it has been reported widely that highly orientated microstructures and strongly coupled grain boundaries (GBs) are required to achieve high critical current density, J$_c$, in polycrystalline forms of the material. The nature of the GBs also determines critically the dependence of J$_c$ on magnetic field. Therefore, a key issue for practical applications of polycrystalline MgB$_2$ is the effect of microstructure and the nature of the GBs on its current carrying ability. In this paper, we report Magneto-Optical (MO) studies of a single-phase polycrystalline MgB$_2$ sample. The results show clear evidence for strongly linked current flow through the GBs over a wide range of temperatures up to $\approx$ 30K. These underline the potential of MgB$_2$ for high performance engineering applications.

Bulk sintered MgB$_2$ samples ($\approx$5mm $\times$ $\approx$5mm $\times$ $\approx$1mm) were prepared from a mixture of high purity Mg and B powders, details of which are described elsewhere [2]. Figure. 1 shows the X-ray diffraction (XRD) pattern of the sample, which confirms it to be non-oriented, single phase MgB$_2$. The superconducting transition temperature, T$_c$, was determined to be $\approx$ 38K via a standard four-point electrical measurement. The measurements for the magnetic field distribution were performed by the MO technique with a liquid He cryostat attached to a conventional *x-y-z* stage of an optical polarizing microscope [3]. An external water-cooled copper coil positioned around the cold stage was used to generate a DC external magnetic field, B$_{ex}$, perpendicular to the sample surface. The principal element in the MO system is the magneto-optical layer (MOL), which exhibits a large Faraday effect. The MOL with dimensions of 5$\times$5 mm used in this study was produced by MAGISTER Ltd., Russia. This was placed in direct contact with the polished surface of the sample. The basis of the MO technique is that an incident beam of polarized light will undergo Faraday rotation when scatterered from a sample surface in the presence of a perpendicular component of magnetic field. After passing through an analyzer, this polarised light provides a direct image of the distribution of B$_z$ near the sample surface.

Figure 2 shows the MO images of the MgB$_2$ sample at B$_{ex}$ = 0 for different temperatures between 5K and 34K. These data were obtained by cooling the sample in a perpendicular external field (FC) of 150mT to 5K. The field was then removed and the sample temperature increased progressively from 5K to 34K. The dark and light regions in the MO images indicate low and high fields, respectively. It can be seen that the magnetic field is trapped uniformly by the whole of the sample at low temperature and that this gradually disappears with increasing temperature. It can be seen further that the magnetic flux escapes preferentially from the sample edge and through weak regions in the sample, which were confirmed by optical microscopy to be cracks. Despite this, the trapped field distributions are relatively uniform up to 30K, indicating that inter-granular current dominates the current flow within the sample. Conversely, island-like distributions of magnetic field are observed in the MO images at 32K. SEM microscopy has been used to confirm the polycrystalline nature of the sample with a typical grain diameter of several tens of microns. The size of the islands of trapped field above 32K correlates remarkably well with this length scale, indicating that the flux distribution is predominantly intra-granular in the vicinity of $T_c$. This suggests further that the intra-granular current dominates the sample properties above 32K and the inter-granular current below it. The very weak MO image at 34K reveals a significant decrease of critical current density as $T_c$ is approached.

Figure 3 shows the field dependent MO images for the same area shown in Fig. 2 after zero-field cooling (ZFC) to (a) 5K and (b) 30K in separate experiments. The external magnetic field of between 30mT and 150mT was again applied perpendicular to the sample surface. Fig. 3, therefore, shows the initial penetration of the magnetic field into the sample. It can be seen that applied fields of up to 150mT are almost completely shielded from the interior of the sample at 5K. This suggests that the grains are extremely well connected in this regime. As a result high currents are able to cross grain boundaries (i.e. to yield a high inter-granular $J_c$). Increasing the temperature to 30K allowed more penetration of magnetic field than at the lower temperature, due primarily to the temperature dependence of the critical current. In this case the magnetic field penetrated initially through the cracks in the sample. However, despite the presence of these weak regions in the specimen, the field distributions elsewhere are relatively uniform with no magnetic field apparent around the sample centre. These data confirm that the sample is strongly coupled between grains (i.e. $J_c$ is dominated by the inter-granular current), even at 30K. In general, good agreement between FC and ZFC behaviour is observed in Figs. 2 and 3. The magnetic field gradient across the regions where flux has penetrated allows $J_c$, to be estimated as $\approx 10^4$ Acm$^{-2}$ at 30K and 150mT. This value agrees well with magnetic measurements of this parameter using SQUID magnetometry on similar material [2]. Larbalestier et al have also reported MO evidence for strongly linked current flow in MgB$_2$ in a multi-phase sample [4]. More homogeneous and symmetrical MO images are obtained for the single-phase sample used in this study, however, which provides direct evidence of inter-granular current flow over a wide range of grain boundary orientations necessarily present in an untextured polycrystalline sample.

In conclusion, MO studies on single-phase polycrystalline MgB$_2$ superconductor have produced direct and clear evidence for high inter-granular current flow through the grain boundaries in this material. The current within the sample is dominated by the inter-granular current up to temperatures of $\approx 30K$ with an estimated $J_c$ of $\approx 10^4$ Acm$^{-2}$ at 30K and 150mT. In contrast to cuprate HTS materials, which require highly oriented microstructures and very small grain boundary angles in order to achieve a good inter-granular connectivity [5], the current flow in MgB$_2$ superconductors appears to be not limited by weakly-linked grain boundaries [4]. The MO images reported here suggest that the electrical properties of polycrystalline MgB$_2$ are isotropic. The observation of high $J_c$ in an untextured microstructure has exciting implications for the potential of this material for a wide range of engineering applications.

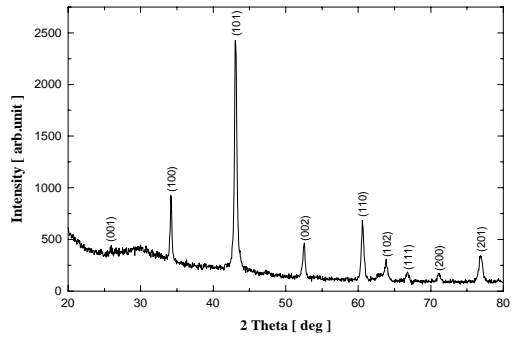

Fig. 1 X-ray diffraction pattern of MgB$_2$ bulk.

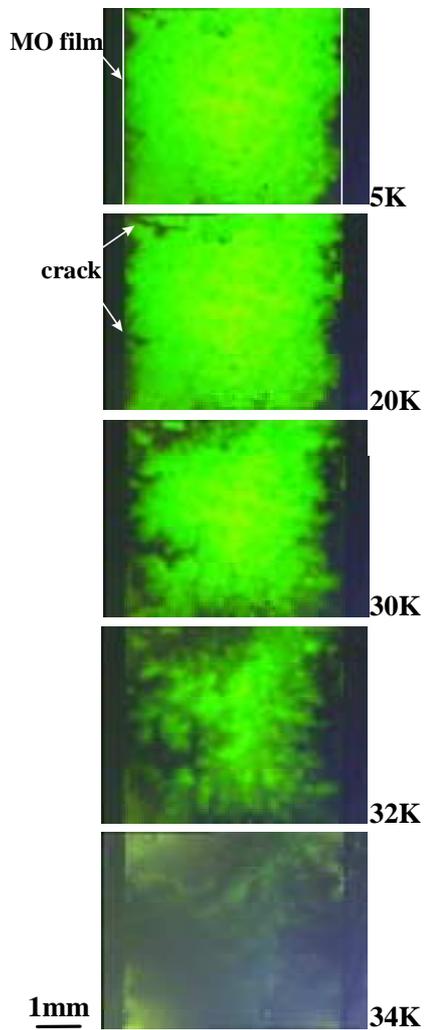

Fig. 2 MO images at 0mT with different temperature from 5K up to 34K after field cooling (FC) process with 150mT of the perpendicular external field.

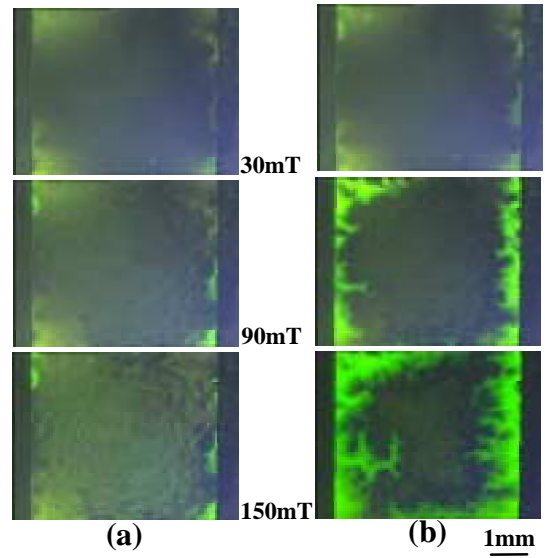

Fig. 3 MO images at 5K: (a) and 30K: (b) with different magnetic fields from 30mT up to 150mT after zero field cooling (ZFC) process. The external magnetic fields were applied perpendicular to sample surface.